\documentclass{aastex}

\newcommand{\msun}{$M_{\odot}\ $}

\shorttitle{Temperature Profiles of Accretion Disks}
\shortauthors{Bhattacharyya et al.}

\begin{document}

\title{Temperature Profiles of Accretion Disks around Rapidly \\ 
Rotating 
Neutron Stars in General Relativity and Implications for Cygnus X-2}

\author{Sudip Bhattacharyya\altaffilmark{1} \and
Arun V. Thampan\altaffilmark{2}}
\affil{Indian Institute of Astrophysics, Bangalore 560 034, INDIA} 

\author{Ranjeev Misra\altaffilmark{3}}
\affil{Inter-University Centre for Astronomy and 
Astrophysics, Pune 411 007, INDIA}

\and

\author{Bhaskar Datta\altaffilmark{4,5}}
\affil{Indian Institute of Astrophysics, Bangalore 560 034, INDIA}

\altaffiltext{1}{Joint Astronomy Program, Indian Institute 
of Science, Bangalore 560 012, INDIA; sudip@iiap.ernet.in;
sudip@physics.iisc.ernet.in}
\altaffiltext{2}{Present Address: Inter--University Centre for
Astronomy and Astrophysics (IUCAA), Pune 411 007, INDIA;
arun@iucaa.ernet.in}
\altaffiltext{3}{Present Address: Department of Physics and Astronomy,
Northwestern University, 2131 Sheridan Road, Evanston,IL 60208--2900;
ranjeev@yawara.astro.nwu.edu}
\altaffiltext{4}{Raman Research Institute, Bangalore 560 080, INDIA}
\altaffiltext{5}{Deceased: Dec. 3, 1999}

\begin{abstract}
We calculate the temperature profiles of (thin) accretion disks around 
rapidly rotating neutron stars (with low surface magnetic fields), taking 
into account 
the full effects of general relativity.  We then consider a model 
for the spectrum of the X--ray emission from the disk, parameterized by
the mass accretion rate, the color temperature and the rotation rate 
of the neutron star.  We derive constraints on these parameters for the
X--ray source Cygnus X--2 using the estimates of the maximum temperature in
the disk along with the disk and boundary layer luminosities, using the 
spectrum inferred from the EXOSAT data.  Our calculations suggest that the 
neutron star in Cygnus X--2 rotates close to the centrifugal 
mass--shed limit. Possible constraints on the neutron star equation of
state are also discussed. 
\end{abstract}

\keywords{X-rays:binaries-X-rays:spectra -stars:neutron -stars:rotation 
-Cygnus X-2}

\section{Introduction} \label{sec: I} 

The soft X--ray spectra of luminous low--mass X--ray binaries (LMXBs) are
believed to originate in geometrically thin accretion disks around neutron
stars with weak surface magnetic fields (see for e.g. White 1995). 
An important parameter in modeling these
spectra is the maximum value of the effective temperature in the accretion 
disk. The effective temperature profile in the disk can be estimated
(assuming the disk to radiate from its surface like a blackbody) if one
knows the accretion energy released in the disk.
In a Newtonian treatment, the innermost region of an accretion disk
surrounding a neutron star with weak magnetic field will extend rather
close to the neutron star surface. The amount of energy released in the
disk will be one--half of the total accretion energy, the other half being
released in the thin boundary layer between the disk's inner edge and the
neutron star's surface. This then gives the disk effective temperature
$(T_{\rm eff})$ varying with the radial distance $(r)$ as 
$T_{\rm eff} \propto r^{-3/4}$
and the maximum effective temperature $(T^{\rm max}_{\rm eff})$ will depend on
the (nonrotating) neutron star mass $(M)$ and radius $(R)$
as $T_{\rm eff}^{\rm max} \propto (M \dot{M}/R^3)^{1/4}$,
where $\dot{M}$ is the steady state mass accretion rate. The value of
$(T^{\rm max}_{\rm eff})$ in the disk, in this approach, occurs at a radial
distance $1.36~R$.  

Mitsuda et al. (1984) parameterized the disk
spectrum by the maximum temperature of the disk, using the above 
formalism and assuming the mass of the neutron star is equal to 
$1.4$~\msun.  These authors assumed that the inner parts of the disk 
do not contribute to the X--ray spectrum, and suggested a multi--color 
spectrum for the X--ray emission from the disk. It was shown by these 
authors, that the observed spectra of Sco X--1, 1608--52, GX 349+2 and 
GX 5--1, 
obtained with the {\it Tenma} satellite, can be well fitted with the 
sum of a multi--color spectrum and a single blackbody spectrum (presumably
coming from the boundary layer). White, Stella \& Parmar (1988) (WSP) 
suggested
that the simple blackbody accretion disk model should be modified
to take into account the effects of electron scattering.
Using {\it EXOSAT} observations, these authors compared the spectral 
properties of the
persistent emission from a number of X--ray burst sources 
with various X--ray emission models.  
This work suggests that either the neutron star (in each system 
considered) rotates close to equilibrium with the Keplerian
disk, or that most of the boundary layer emission is not represented
by a blackbody spectrum.  

For accretion disks around compact objects, one possibility is that of 
the accretion disk not being Keplerian in nature.  For e.g. Titarchuk, 
Lapidus \& Muslimov (1998) have formulated a boundary problem in which 
the Keplerian accretion flow in the inner disk is smoothly adjusted 
to the neutron star rotation rate.  The generality of such a formulation 
permits application even to black holes, but only for certain assumed inner 
boundary conditions.  These authors demonstrate that there exists a 
transition layer (having an extent of the order of the neutron star radius) 
in which the accretion flow is sub-Keplerian.  An attractive feature of this 
formalism is that it allows super-Keplerian motion at the outer boundary of 
the transition layer, permitting the formation of a hot blob that ultimately 
bounces out to the magnetosphere. This formalism (Titarchuk \& Osherovich 
1999; Osherovich \& Titarchuk 1999a; Osherovich \& Titarchuk 1999b; 
Titarchuk, Osherovich \& Kuznetsov 1999) therefore provides a mechanism 
for the production of high frequency quasi--periodic oscillations (QPOs) 
observed in the X--ray flux from several LMXBs.  Such effects, when taken
into account, can modify the Newtonian disk temperature profile 
(Chakrabarti \& Titarchuk 1995).

There are several other effects which will modify the Newtonian disk 
temperature profile, such as the effects of general relativity and of 
irradiation of the disk by the central neutron star. The wind mass loss 
from the disk and the residual magnetic field near the disk's inner edge 
may also play a part in modifying the effective temperature (Knigge 1999). 
Czerny, Czerny \& Grindlay (1986) calculated LMXB disk spectra assuming 
that a disk radiates locally as a blackbody with the energy flux detemined 
by viscous forces, as well as irradiation by the boundary layer, and took 
into account relativistic effects, some of them in an approximate way.
The possible effects of general relativity were also discussed by 
Hanawa (1989), using the Schwarzschild (nonrotating) metric, assuming that 
the neutron star radius is
less than the radius of the innermost stable circular orbit 
($r_{\rm in} = 6 G M/c^2$), which they identified 
as the disk inner boundary. The color temperature
was assumed to be higher than the effective temperature by a factor
of 1.5.  It was found by Hanawa (1989) that the observations are
consistent with a geometrically thin, optically thick accretion 
disk, whose inner edge is at $r=r_{\rm in}$, $r$ being the Schwarzschild
radial coordinate. 

An important dynamical aspect of disk accretion on to a weakly magnetized 
neutron star is that the neutron star will get spun up to its equilibrium 
period, which is of the order of milliseconds (see Bhattacharya \& van den 
Heuvel 1991, and refereces therein).
The effect of rotation is to increase the equatorial radius of the
neutron star, and also to relocate the innermost stable circular orbit (for a 
corotating disk) closer
to the stellar surface (as compared to the Schwarzschild case).
These effects will be substantial for rapid rotation rates in a fully
general relativistic treatment that includes rotation. Therefore, for 
accreting neutron stars with low magnetic fields,  the
stellar radius can be greater or less than the radius of the
innermost stable orbit, depending on the neutron star equation of state
and the spacetime geometry.  The effect of magnetic field will be to 
constrain the location of the inner--edge of the accretion disk to the 
magnetospheric (Alf\'{v}en) radius. In such a case, $r_{\rm in}$ would lose 
the astrophysical relevance as discussed here.  However, this will be so 
only if the magnetic field strength ($B$) is large.  The problem addressed
in this paper refer to LMXBs which contain old neutron stars which are
believed to have undergone sufficient magnetic field decay (Bhattacharya \&
Datta 1996).  Clearly, for low magnetic field case, a number of different 
disk geometries will be possible if general relativistic effects of rotation 
are taken into account. These structural differences influence the effective 
temperature profile and the conclusions derived by Czerny, Czerny \& Grindlay 
(1986) and Hanawa (1989) are likely to be modified.  

In this paper, we attempt to highlight the effects brought in due to general 
relativity and rotation of the neutron star on the accretion disk temperature 
profile and then 
apply this to the particular case of the X--ray source Cygnus X--2.  For 
simplicity (unlike Titarchuk, Lapidus \& Muslimov 1998), we assume the 
accretion disk to be fully Keplerian, geometrically thin and optically thick.  
We first give a theoretical estimate of the modifications in 
$T_{\rm col}^{max}$ that would result if inclusion is made of the rotational
effects of general relativity, and illustrate these by taking representative
neutron star equations of state. We then consider
a model for the spectrum parameterized by the mass accretion rate,
the color factor, and the rotation rate of the accreting 
neutron star (assumed to be weakly magnetized).  We derive constraints on 
these parameters for the X--ray source 
Cygnus X--2, for which we take the estimates of $T_{\rm eff}^{\rm max}$,
the disk luminosity and boundary layer luminosity from the analysis
of WSP.  A conclusion of our work is that
the neutron star in Cygnus X--2 has a rapid spin rate
 close to the centrifugal mass--shed limit. 

The format of this paper is as follows. In Section~\ref{sec: II}, we
discuss the rotational general relativistic effects on the disk temperature,
using a formalism given by Page \& Thorne (1974) and the disk irradiation 
by the neutron star.  The theoretical 
predictions  for the temperature profiles with these effects taken into 
account are presented in Section~\ref{sec: III}. Section~\ref{sec: IV} 
deals with comparison with
observations, and its implications for parameters of our model for
Cygnus X--2. A summary and discussions are presented in Section~\ref{sec: V}.

% Section II   THE EFFECTIVE TEMPERATURE OF THE DISK
\section{The Effective Temperature of the Disk}\label{sec: II} 

%\begin{center}
\subsection{Effects of General Relativity and Rotation}
%\end{center}

The effective temperature in the disk (assumed to be optically thick) is
given by
% equation (1)
\begin{eqnarray}
T_{\rm eff} & = & (F/\sigma)^{1/4} \label{eq: teff}
\end{eqnarray}
\noindent where $\sigma$ is the Stephan--Boltzmann constant and $F$ is
the X--ray energy flux per unit surface area. We use the formalism
given by Page \& Thorne (1974), who gave the following general relativistic 
expression for $F$ emitted from the surface of an (geometrically thin and non--
self--gravitating) accretion disk around a rotating black hole:
% equation (2)
\begin{eqnarray}
F(r) & = & \frac{\dot{M}}{4 \pi r} f(r)
\end{eqnarray}
\noindent where
% equation (3)
\begin{eqnarray}
f(r) & = & -\Omega_{{\rm K},r} (\tilde{E} - \Omega_{\rm K} \tilde{l})^{-2}
\int_{r_{\rm in}}^{r} (\tilde{E} - \Omega_{\rm K} \tilde{l}) \tilde{l}_{,r} dr
\label{eq: fr}
\end{eqnarray}
\noindent Here $r_{\rm in}$ is the disk inner edge radius, $\tilde{E}$, 
$\tilde{l}$
are the specific energy and specific angular momentum of a test particle
in a Keplerian orbit and $\Omega_{\rm K}$ is the Keplerian angular velocity at
radial distance $r$. In our notation, a comma followed by a variable as
subscript to a quantity, represents
a derivative of the quantity with respect to the variable. Also, in this 
paper, we use the geometric units $c=G=1$.

For accreting neutron stars located within the disk inner edge,
the situation is analogous to the black hole binary case, and the 
above formula, using a metric describing a rotating neutron
star, can be applied directly for our purpose. However, unlike the black 
hole binary case, there can be situations for neutron star binaries
where the  neutron star
radius exceeds the innermost stable circular orbit radius.
In such situations, the boundary condition, assumed by Page \& Thorne
(1974),
 that the torque vanishes 
at the disk inner edge will not be strictly valid.
Use of Eq. (\ref{eq: teff}) will then be an approximation. This 
will affect the temperatures close to the disk inner edge, but not 
the $T_{\rm eff}^{\rm max}$ to any significant degree (see section 5 
for discussion).  
% comments on approximate validity of the equation in such cases

In order to evaluate $T_{\rm eff}$ using Eq. (\ref{eq: teff}), we need to 
know the radial profiles of $\tilde{E}$, $\tilde{l}$ and $\Omega_{\rm K}$.  For 
this, we have to first compute the equilibrium sequences of neutron stars in 
rapid rotation.  These can be calculated by noting that the space--time around 
a rotating neutron star can be described by the following metric
(Cook, Shapiro \& Teukolsky 1994):
\begin{eqnarray}
ds^{2} & = & g_{\lambda \beta} dx^{\lambda} dx^{\beta} 
        ~~~(\lambda, \beta = 0,1,2,3) \nonumber \\
       & = & - e^{\gamma + \rho} dt^{2}
             + e^{2 \alpha} (d\bar{r}^{2} + \bar{r}^{2} d \theta^{2})
             + e^{\gamma - \rho} \bar{r}^{2} sin^{2} \theta (d \phi - 
            \omega dt)^{2} ~,
\label{eq: metric}
\end{eqnarray}
%\vspace{-1.0cm}
\noindent where the metric potentials $\gamma$, $\rho$, $\alpha$, and the 
angular velocity ($\omega$) of the stellar fluid relative to the local 
inertial frame are all functions of the quasi--isotropic radial coordinate 
($\bar{r}$) and polar angle ($\theta$).  
$\bar{r}$ is related to the Schwarzschild radial coordinate ($r$) through the 
equation $r= \bar{r}e^{(\gamma-\rho)/2}$.  

On the assumptions that the matter is a perfect fluid and that the 
space--time described by metric (\ref{eq: metric})  is stationary, 
axisymmetric, 
asymptotically flat and reflection--symmetric (about the equatorial 
plane), the
Einstein field equations reduce to three non--homogeneous, 
second--order, coupled
differential equations (for $\gamma$, $\rho$ and $\omega$) and one ordinary
differential equation (for $\alpha$) in terms of $\epsilon$ and $P$ (the total
energy density and the pressure of neutron star matter respectively) in 
addition to terms involving $\gamma$, $\rho$, $\omega$ and $\alpha$ 
(see Komatsu, Eriguchi \& Hachisu 1989). We have solved these equations
(self--consistently and numerically) to obtain $\gamma$, $\rho$, $\omega$, 
$\alpha$, $P$ and $\Omega$ (which is the angular velocity of the neutron star 
matter as measured by a distant observer) as functions of $\bar{r}$ and 
$\theta$. The angular velocity enters in the 
equations through the rotation law (which must be specified) for the matter 
distribution.  The equilibrium solutions so obtained can then be used to 
calculate bulk structure parameters such as gravitational mass $M$, 
equatorial radius $R$, angular momentum $J$, etc. of the rotating neutron 
star. We assume that the neutron star rotates rigidly. Thus, $\Omega$ is 
constant for the stellar matter distribution, and is taken to be equal to
$\Omega_{\ast}$, defined as the angular velocity of the neutron star as 
measured by a distant observer.  

Eq. (\ref{eq: teff})  gives the effective disk temperature $T_{\rm eff}$ 
with respect to an observer comoving with the disk.  From the observational
viewpoint this temperature must be modified, taking into account the 
gravitational redshift and the rotational Doppler effect.
%For a general stationary axisymmetric metric, this is (see Luminet 1979):
%\begin{eqnarray}
%1+z & = & (1 + \Omega_{\rm K} b \sin{i} \sin{\alpha} ) 
%(-g_{tt} - 2\Omega_{\rm K} g_{t \phi} - \Omega_{\rm K}^{2} g_{\phi \phi})^{-1/2}
%\label{eq: redshift}
%\end{eqnarray}
%\noindent where $b$ is the photon impact parameter, $i$ is the inclination
%angle and $\alpha$ is a function of $i$ and the azimuthal angle and 
%$\Omega_{\rm K}$ is the Keplerian angular velocity (there is a minor 
%error in Luminet's paper Eq. (18); $\Omega_{\rm K}$ and $\Omega_{\rm K}^2$
%multiplying the metric elements $g_{t \phi}$ and $g_{\phi \phi}$ are 
%missing).
In order to keep our analysis tractable, we use 
%Eq. (\ref{eq: redshift}) corresponding to the Schwarzschild
%value (with zero inclination angle):
the expression given in Hanawa (1989) for this modification : 
\begin{eqnarray}
1+z = (1-\frac{3M}{r})^{-1/2}
\end{eqnarray}
%\noindent (see Hanawa 1989).  
With this correction for $(1+z)$, we define a temperature relevant for
observations ($T_{\rm obs}$) as:
\begin{eqnarray}
T_{\rm obs} = \frac{1}{1+z} T_{\rm eff} \label{eq: tobs}
\end{eqnarray}

%\begin{center}
\subsection{Computation of $\tilde{E}$, $\tilde{l}$ and $\Omega_{\rm K}$}
%\end{center}

For the work presented in this paper, we compute constant gravitational mass 
($M$) equilibrium sequences for rigidly and rapidly rotating 
neutron stars using the formalism described above (see Datta, Thampan \& 
Bombaci 1998 for details), keeping in mind the importance of the parameters
$M$ and $\Omega_{\ast}$ for modeling the X--ray emission from LMXBs.
These sequences are constructed starting from 
the static limit all the way upto the rotation rate corresponding to the 
centrifugal
 mass  shed limit. The latter limit corresponds to the maximum 
$\Omega_{\ast}$ (=$\Omega_{\rm ms}$)
for which  centrifugal forces are able to  balance the inward gravitational 
force.  We now briefly describe how the quantities $\tilde{E}$, $\tilde{l}$
and $\Omega_{\rm K}$ are calculated; for details, the reader is referred
to Thampan \& Datta (1998).  For a material particle in the gravitational 
field described by metric~(\ref{eq: metric}), we can write down 
(see for e.g. Misner, Thorne \& Wheeler 1973) the equation of motion in 
the equatorial plane.  These will be in terms of $\tilde{E}$, $\tilde{l}$,
$\omega$, $\bar{r}$ and the metric coefficients.  The equation of motion in the 
radial direction defines the effective gravitational potential.  The 
two conditions for orbits (circularity and extremum) at any $r$ yield
values for $\tilde{E}$ and $\tilde{l}$ as given by:
\begin{eqnarray}
\tilde{E} - \omega \tilde{l} & = & \frac{e^{(\gamma+\rho)/2}}{\sqrt{1-v^2}} \\
\tilde{l} & = & \frac{v \bar{r} e^{(\gamma-\rho)/2}}{\sqrt{1-v^2}}
\end{eqnarray}
\noindent where $v=(\Omega-\omega) \bar{r} e^{-\rho} sin \theta$ is the 
physical
velocity of the matter.  The equation of motion in the azimuthal direction
and that in the time direction yield the Keplerian angular velocity as
\begin{eqnarray}
\Omega_{\rm K} & = & e^{2\rho(\bar{r})}
\frac{\tilde{l}/ \bar{r}^2}{(\tilde{E}-\omega \tilde{l})} + \omega(\bar{r})
\end{eqnarray}

%We then compute, for a co--rotating disk, the 
%profiles for $\tilde{E}$, $\tilde{l}$, $\Omega_{\rm K}$, and the quantities 
%$E_{\rm BL}$ and $E_{\rm D}$, defined as the specific gravitational energy
%released in the boundary layer and in the disk repectively (see
%Thampan \& Datta 1998). 

%\begin{center}
\subsection{Computation of $E_{\rm BL}$ and $E_{\rm D}$}
%\end{center}

We define the specific gravitational energy release due to the ingress of a 
material particle from infinity to the disk inner edge as $E_{\rm D}$, and 
that due to the particle spiralling in from the disk inner edge to the surface 
of the star as the boundary layer energy: $E_{\rm BL}$.  For the case where
the disk inner edge coincides with the stellar surface, $E_{\rm BL}$ is
the difference in the energy of the particle in a Keplerian orbit 
at $r=R$ and that when it is at rest on the stellar surface.  The
exact expressions for $E_{\rm BL}$ and $E_{\rm D}$ are determined
by the effective potential corresponding to any given space--time 
metric. For the Schwarzschild metric and the `slow'--rotation 
Hartle--Thorne metric, the boundary layer to disk luminosity 
ratio has been calculated by Sunyaev \& Shakura (1986) and Datta,
Thampan \& Wiita (1995) respectively.
Calculations of $E_{\rm BL}$ and $E_{\rm D}$ corresponding 
to the metric (\ref{eq: metric}) and used for the modeling in this
paper are discussed in detail in Thampan \& Datta (1998).

% subsection II.I   THE EFFECT OF IRRADIATION
%\subsection{The Effect of Irradiation}
%\begin{center}
\subsection{Disk Irradiation by the Neutron Star}
%\end{center}

For luminous LMXBs, there can be substantial irradiation of the disk surface
by the radiation coming from the  neutron star boundary layer.  The radiation
temperature at the surface of a disk irradiated by a central source is given
by  (King, Kolb \& Burderi 1996)
\begin{eqnarray}
T_{\rm irr}(r) & = & 
\left(\frac{\eta \dot{M}c^2(1-\beta)}{4\pi \sigma r^2} \frac{h}{r} (n-1) \right)^{1/4}
\label{eq: tirr}
\end{eqnarray}
\noindent where $\eta$ is the efficiency of conversion of accreted rest mass
to energy, $\beta$ is the X--ray albedo, $h$ is the 
half--thickness of the disk at $r$ and $n$ is given by the relation
$h \propto r^n$.  For actual values of $\beta$, $h/r$ and $n$, needed for
our computation here, we choose the same values (i.e., 0.9, 0.2 and 9/7 respectively) 
as given in King, Kolb 
\& Burderi (1996). Although the above equation is derived based on Newtonian
considerations, corrections due to general relativity (including that of 
rapid rotation) will be manifested through the factor $\eta$.  We have made a 
general relativistic evaluation
of $\eta$ for various neutron star rotating configurations, corresponding
to realistic neutron star EOS models, as described in Thampan \& Datta (1998).
Since $T_{\rm irr}(r) \propto r^{-1/2}$ and 
$T_{\rm eff}(r) \propto r^{-3/4}$,   $T_{\rm irr}$ will dominate over 
$T_{\rm eff}$ only at large distances.  
The net effective
temperature of the disk will be given by  (see Vrtilek et al. 1990)
\begin{eqnarray}
T_{\rm disk}(r) & = & (T_{\rm eff}^4(r) + T_{\rm irr}^4(r))^{1/4}
\label{eq: tdisk}
\end{eqnarray}
For the Cygnus X--2 modeling presented here, we find that $T_{\rm irr}$ 
does not play any significant role.  However, since this quantity has 
consequences for the disk instability, we calculate it using 
Eq.~(\ref{eq: tirr}), and illustrate it for the rotating neutron star
models considered here.
%This has consequences for the disk 
%stability, and is briefly discussed in the last section. 
% Section III 
\section{Results for the Disk Temperature Profile}\label{sec: III}

%\subsection{Neutron star equations of state}
%\begin{center}
\subsection{Neutron Star Equations of State}
%\end{center}

The neutron star EOS is an important determining factor 
for the structure parameters of the star.  A variety of neutron star EOS is 
available in the literature, ranging from very soft to very stiff models.  For 
the purpose of our calculation, we have chosen four EOS models: 
(A) Pandharipande (1979) (hyperons), (B) Baldo, Bombaci \& Burgio (1997)
(AV14 + 3bf), (C) Walecka (1974) and (D) Sahu, Basu \& Datta (1993).  
Of these, model (A) is soft, (B) is intermediate in stiffness and 
(C) and (D) are stiff EOS.  With this representative choice of EOS, 
the results of our calculations are expected to be of sufficient generality.

%\subsection{The Results}
%\begin{center}
\subsection{The Results}
%\end{center}

We have calculated the disk temperature profiles for rapidly rotating, 
constant gravitational mass sequences of neutron stars in general relativity. 
For our purpose here, we choose two values for the gravitational mass, namely, 
$1.4$~\msun and $1.78$~\msun, the former being the canonical mass for neutron 
stars (as inferred from binary X--ray pulsar data), while the latter is the 
estimated mass for the neutron star in Cygnus X--2 (Orosz \& Kuulkers 1999).
It may be noted with caution (Haberl \& Titarchuk 1995), that this value
is not confirmed from X--ray burst spectral analysis.  We use the value of 
$M=1.78$~\msun for the illustration of our results, and leave the issue for
future confirmation.  In order to make a comparison with observations of 
Cygnus X--2, we  need to calculate the values of the 
$E_{\rm BL}$ and $E_{\rm D}$, and $T_{\rm eff}^{\rm max}$
as functions of the stellar rotation rate ($\Omega_{\ast}$),
for the above chosen values of the gravitational mass ($M$). 

In Table 1, we list the values of the stellar rotation rate at centrifugal 
mass shed limit ($\Omega_{\rm ms}$), the neutron star radius ($R$), 
the radius of the inner edge of the disk ($r_{\rm in}$), $E_{\rm BL}$, 
$E_{\rm D}$ and the ratio $E_{\rm BL}/E_{\rm D}$, $T_{\rm eff}^{\rm max}$ 
 \& $T_{\rm obs}^{\rm max}$, $r_{\rm eff}^{\rm max}$
 and $r_{\rm obs}^{\rm max}$, for the two mentioned values of 
$M$ and for the different EOS models. 
The last nine computed quantities 
are given for two values of neutron star rotation rate, namely, 
the static limit ($\Omega_{\ast}=0$) and the centrifugal mass shed 
limit ($\Omega_{\ast}=\Omega_{\rm ms}$).  $E_{\rm D}$ and $E_{rm BL}$
are in specific units (i.e. units of rest energy $m_0 c^2$, of the 
accreted partilce).
The temperatures are expressed in units of $\dot{M}_{17}^{1/4}\times 10^5$~K
(where $\dot{M}_{17} = \dot{M}/10^{17}$~g~s$^{-1}$).  
%For example, for a value of $\dot{M}=1.4\times10^{17}$~($M/$\msun)~g~s$^{-1}$
%(the Eddington accretion rate), 
%with $M=1.4$\msun and $T_{\rm obs}^{\rm max}=40.0$ in the 
%table, we have actually $T_{\rm obs}^{\rm max}=4.733\times10^6$~K.  
From this Table it may be seen
that for a given neutron star gravitational mass ($M$):
(a) $\Omega_{\rm ms}$ decreases for increasing stiffness of the EOS model.
(b) $R$ is greater for stiffer EOS.
(c) The behaviour of $r_{\rm in}$ depends on whether $r_{\rm ms}>R$ or 
    $r_{\rm ms}<R$ and hence appears non--monotonic.
(d) $E_{\rm BL}$ for the non--rotating configuration decreases with stiffness
    of the EOS.  For a configuration rotating at the mass shed limit, 
    $E_{\rm BL}$ is insignificant.
(e) In the non--rotating limit, $E_{\rm D}$ remains roughly constant for
    varying stiffness of the EOS model.  However, for the rapidly rotating
    case, the value of $E_{\rm D}$ decreases with increasing stiffness.
(f) The ratio $E_{\rm BL}/E_{\rm D}$ in static limit is highest for the
    softest EOS model.  For the rapidly rotating case, this ratio is 
    uniformly insignificant.
(g) $T_{\rm eff}^{\rm max}$ and $T_{\rm obs}^{\rm max}$ decrease with 
    increasing stiffness
    of the EOS  models.  However, these values exhibit non--monotic variation
    with $\Omega_{\ast}$ (see Fig. 5 for the first parameter).
(h) The rest of the parameters, namely, $r_{\rm eff}^{\rm max}$ and 
     $r_{\rm obs}^{\rm max}$ are non--monotonic with respect to 
     the EOS stiffness parameter.

In Fig. 1, we display the variation of $R$ (the dashed curve) and 
$r_{\rm in}$ (the continuous curve) with 
$\Omega_{\ast}$ for $M=1.4$~\msun for the four EOS models that we
have chosen.  From this figure it is seen that for a constant gravitational 
mass sequence, for both soft and intermediate EOS models, $r_{\rm in} > R$ for 
slow rotation rates whereas, for rapid rotation rates $r_{\rm in} = R$. 
In other words, for neutron stars spinning very rapidly, the inner edge of
the disk will almost coincide with the stellar surface. It may be noted that 
for the stiff EOS models, this condition obtains even at slow rotation
rates of the neutron star.  
%This qualitative feature has a significant
%effect on the temperature profiles as can be seen in Fig. 3a.

It is instructive to make a comparison of the temperature profiles 
calculated using a Newtonian prescription with that
obtained in a relativistic description using Schwarzschild metric.
This is shown in Fig. 2, for the EOS model (B) and $M=1.4$~\msun 
(the trend is similar for all the EOS).  The vertical axis in 
this figure is $T_{\rm eff}$ (in this and all other figures, the
temperatures are shown in units of $\dot{M}_{17}^{1/4}$) and  the
horizontal axis, the radial distance in km.
This figure underlines the importance of general relativity in determining
the accretion disk temperature profiles; 
the Schwarzschild result for $T_{\rm eff}^{\rm max}$ is always less than
the Newtonian result, and 
for the neutron star configuration considered here, the overestimate
is almost $25$\%. 
For the sake of illustration, we also show the corresponding 
curve for a neutron star rotating at the mass shed limit (curve 4, Fig. 5).
The disk inner edge is at the radius of the 
innermost stable circular orbit  for all the cases.
%(in the Newtonian case, we terminate 
%the temperature profile at $r_{\rm in}=6GM/c^2$).
Note that the disk inner edge should be at $R$ for Newtonian case; but we
 have taken $r_{\rm in}=6GM/c^2$ as assumed in Shapiro \& Teukolsky (1983).  

The effect of neutron star rotation on the accretion disk temperature, treated 
general relativistically, is illustrated in Fig. 3a and 3b.  
Fig. 3a corresponds to the EOS model (B). The qualitative features 
of this graph are similar for the other EOS models, and are not
shown here.
However, the temperature profiles exhibit a marked dependence on 
the EOS.  This dependence is illustrated in Fig. 3b, which is 
done for a particular value of $\Omega_{\ast}= \Omega_{\rm ms}$.
All these temperature profiles have been calculated for a neutron star
mass equal to $1.4$~\msun.   
The temperature profiles shown in Fig. 3a do not have a monotonic behavior 
with respect to $\Omega_{\ast}$.  This behavior is a composite of two 
underlying
effects: (i) the energy flux emitted from the disk increases with 
$\Omega_{\ast}$ and (ii) the nature of the dependence of $r_{\rm in}$ 
(where $T_{\rm eff}$ vanishes : the boundary condition)
on $\Omega_{\ast}$ (see Fig. 1). This is more clearly brought out in
Fig. 4, where we have plotted of 
$T_{\rm eff}$ versus $\Omega_{\ast}$ for selected 
constant radial distances (indicated in six different panels) and EOS (B). 
At large radial distances, the value $T_{\rm eff}$
is almost independent of the boundary condition; hence the 
temperature always increases with $\Omega_{\ast}$ in Fig. 4 f.

The variations of $E_{\rm D}$, $E_{\rm BL}$, the ratio 
$E_{\rm BL}/E_{\rm D}$ and 
$T_{\rm eff}^{\rm max}$ with $\Omega_{\ast}$ 
are displayed in Fig. 5 for all EOS models considered here.  All the plots
correspond to $M=1.4$~\msun.  Unlike constant central density neutron star 
sequences (Thampan \& Datta 1998), for the constant gravitational mass 
sequences, $E_{\rm D}$ does not have a general monotonic behaviour with 
$\Omega_{\ast}$.  $T_{\rm eff}^{\rm max}$ has a behaviour akin 
to that of $E_{\rm D}$ (because of the reasons mentioned earlier).  
$E_{\rm BL}$ decreases with $\Omega_{\ast}$, slowly at first but
rapidly as $\Omega_{\ast}$ tends to $\Omega_{\rm ms}$. The variation of 
$E_{\rm BL}/E_{\rm D}$ with respect to $\Omega_{\ast}$ is similar to 
that of $E_{\rm BL}$.

We provide a comparison between the effective temperature
(Eq. \ref{eq: teff}) and the irradiation temperature (Eq. \ref{eq: tirr}),
in Fig. 6. We have taken $\eta=E_{\rm BL}+E_{\rm D}$.  Fig. 6a is for 
$\Omega_{\ast}=0$ while 
Fig. 6b is for a higher $\Omega_{\ast} = 6420$ rad s$^{-1}$.  The curves are 
for the gravitational mass corresponding to $1.4$~\msun for the EOS model (B).
The irradiation temperature 
becomes larger than the effective temperature at some large value of 
the radial distance, the ratio of the former to the latter becoming
increasingly large beyond this distance. For $E_{\rm BL}$ small
compared to $E_{\rm D}$ (as will be the case for a rapid neutron 
star spin rate), irradiation effects in the inner disk region will 
not be significant.
  Defining the radial point where
the irradiation temperature profile crosses the effective temperature 
profile as $r=r_{\rm cross}$ and the corresponding temperature as
$T_{\rm cross}$, we display plots of $r_{\rm cross}$ and $T_{\rm cross}$
with $\Omega_{\ast}$ respectively in Figs. (7 a) and (7 b).
%Here the irradiation temperatures are calculated for $\eta=E_{\rm BL}+
%E_{\rm D}$, as $r_{\rm cross}$ in each of these cases is far from the 
%neutron star surface.
It can be seen that $r_{\rm cross}$ increases with $\Omega_{\ast}$, 
just as $E_{\rm S}$ does, and hence the irradiation effect decreases
with increasing $\Omega_{\ast}$. Therefore $T_{\rm cross}$ 
decreases with increasing $\Omega_{\ast}$. 

In Fig. 8, we illustrate the disk temperature 
($T_{\rm disk}$) profile for EOS model (B) corresponding to 
$M=1.4$~\msun for various values of $\Omega_{\ast}$. 
We illustrate the variation of $T_{\rm disk}$ with $\Omega_{\ast}$ at
fixed radial points in the disk in Fig. 9.  The effect of $T_{\rm irr}$ 
on $T_{\rm disk}$ can be noted in Fig. 9f.

\section{Comparison with Observations: Implications for
Cygnus X--2}\label{sec: IV}

The X-ray spectrum will have two contributions: one
from the optically thick disk and the other from the boundary layer
near the neutron star surface. The spectral shape of the disk emission depends
on the accretion rate. For $\dot {M} << 10^{17}$~g~s$^{-1}$, the opacity in 
the disk is dominated
by free-free absorption and the spectrum will be a sum of blackbody spectra.
The temperature of the local spectra (with respect to a co--moving observer) 
will be equal to the temperature $T_{\rm eff}(r)$ at that radius.
The observer at large distance will see a temperature $T_{\rm obs}(r)$,
which includes the effect of gravitational redshift and Doppler broadening,
as mentioned in Section~\ref{sec: II}.
%($T_{\rm obs}$ includes the effect of gravitational redshift and Doppler 
%broadening, as mentioned in section 2).
At higher accretion rates ($\dot {M} \approx 10^{17}$~g~s$^{-1}$) 
the opacity will be dominated by Thomson scattering and the spectrum from 
the disk is that of a modified blackbody (Shakura \& Sunyaev 1973).
However, for still higher accretion 
rates Comptonization in the upper layer of the disk becomes important 
leading to saturation in the local spectrum to form a Wien peak.
The emergent spectrum can then be described as a sum of blackbody
emissions but at a different temperature than $T_{\rm obs}$. The 
spectral temperature seen by a distant observer is the color temperature 
$T_{\rm col}$. In general
$T_{\rm col} = f(r) T_{\rm obs}$ where the function $f$ is called 
the color factor
(or the spectral hardening factor), and it depends on the vertical 
structure of the disk. Shimura \& Takahara (1988) calculated the color 
factor for various accretion rates and masses of the accreting compact
object (black hole) and found that $f\approx$~($1.8$--$2.0$) is nearly 
independent of accretion rate and radial distance, for 
$\dot{M}\sim\dot{M}_{\rm Edd}$.  These authors find that for accretion rate 
$\sim 10$\% of the Eddington limit, $f\approx1.7$.  More recently, however, 
from analysis of the high--energy radiation from GRO J1655-40, a black--hole 
transient source observed by RXTE, Borozdin et al. (1999) obtain a value of 
$f=2.6$, which is higher than previous estimates used in the literature.
With this approximation for $T_{\rm col}$, the spectrum from optically thick 
disks with high accretion rates can be represented as a sum of diluted 
blackbodies.  The local flux at each radius is
\begin{equation}
F_\nu = \frac{1}{f^4} \pi B_\nu (fT_{\rm eff})
\end{equation}
where $B_\nu$ is the Planck function. 
For high accretion rates the boundary layer at the neutron star surface
is expected to be optically thick and an additional single component blackbody
spectrum should be observed.

The {\it EXOSAT} observations of Cygnus X-2 (Hasinger et al. 1986) have been 
fitted to several models by WSP. One of 
the models is a blackbody emission upto the innermost stable circular 
orbit of the
accretion disk and an additional blackbody
spectrum to account for the boundary layer emission. The
spectrum from such a disk is the sum of blackbody emission with a
temperature profile 
\begin{equation}
T \propto r^{-3/4} (1 - (r_{\rm in}/r)^{1/2})^{1/4}
\end{equation}
%where $R$ is the radius and $r_{\rm in}$ is the radius of the innermost stable
%orbit or the neutron star surface. 
WSP have identified this temperature
as the effective temperature which, as mentioned by them, is inconsistent
since the accretion rate for Cygnus X-2 is high ($\dot {M} \approx 
10^{18}$~g~s$^{-1}$). However, as mentioned above, identifying this 
temperature profile as the 
color temperature makes the model consistent if the color factor
is nearly independent of radius. Moreover, the inferred 
temperature profile (i.e. $T_{\rm obs} = T_{\rm col}/f$)  
is similar to the one developed in previous section. Therefore, in this paper
we assume that the maximum of the best fit color temperature profile 
$T_{\rm col}^{\rm max}$
is related to the maximum temperature $T_{\rm obs}^{\rm max}$ computed in previous section
by ($T_{\rm col}^{\rm max} \approx f T_{\rm obs}^{\rm max}$).  
Shimura \& Takahara (1988) 
suggested a value of $1.85$ for the factor $f$, for an assumed neutron
star mass equal to $1.4$~\msun and $\dot{M}=10 \dot{M}_{\rm Edd}$,
where $\dot{M}_{\rm Edd}$ is the Eddington luminosity, with the mass to 
energy conversion efficiency taken as unity. 

For the source Cygnus X--2, the best spectral fit to the data is when
$T_{\rm col}^{\rm max} = 1.8 \times 10^7 $ K, $L_{\rm D} = 2.1 \times 
10^{38}$~ergs~s$^{-1}$ 
and $ L_{\rm BL} = 2.8 \times 10^{37} $~ergs~s$^{-1}$ (WSP), 
where $L_{\rm D}$ is the
disk luminosity and $L_{\rm BL}$, the luminosity in the boundary layer.
The distance to the source  and the inclination angle to
the source have been estimated by Orosz \& Kuulkers (1998) 
to be $\approx 8$ kpc and $60^o$ 
respectively. 
From these values one can obtain, using the formalism described in section 2,
 the angular velocity of the neutron star
($\Omega_{\rm *}$) for a given neutron star mass, accretion rate ($\dot {M}$), 
color
factor $f$ and equation of state. However, in order to make allowance for 
the uncertainties in the fitting procedure and in the value of $z$, and 
also those arising due to the simplicity of the model, we consider a range of 
acceptable
values for $T_{\rm col}^{\rm max}$, $L_{\rm D}$ and $L_{\rm BL}$. 
In particular, we allow for deviations in $T_{\rm col}^{\rm max}$ and
the best fit luminosities:  we take two combinations of these, namely,
($10$\%, $25$\%) and ($20$\%, $50$\%), where the first number in parentheses
corresponds to the error in $T_{\rm col}^{\rm max}$ and the second to
the error in the best fit luminosities.
Note that we neglect the irradiation temperature here, 
as $T_{\rm disk} \approx T_{\rm eff}$ at the inner region of the disk 
(the region where the disk temperature reaches a maximum). 
The constraints on $\dot{M}$, $\Omega_{\ast}$, and $f$ are 
calculated 
for two values of the mass of the neutron star in Cygnus X--2, namely,
$1.4 M_\odot$ and $1.78 M_\odot$.
We obtain a range of consistent values for $\dot {M}$,
$\Omega_{\rm *}$ and $f$ (and hence, allowed ranges of different quantities). 
The procedure is as follows.

As described in previous section, we can calculate
the different quantities ($E_{\rm D}$, $E_{\rm BL}$, 
$T_{\rm obs}^{\rm max}$, $R$, $r_{\rm in}$, etc.) as a function of
$\Omega_{\rm *}$.
Taking the observed (or fitted) values for $T_{\rm col}^{\rm max}$, 
$L_{\rm BL}$ and ($L_{\rm BL}+L_{\rm D}$) with the error bars, 
we have two limiting
values for each of these quantities. We assume a particular
value for each of $f$ and $\dot {M}$, from which we obtain the corresponding
fitted values of $T_{\rm obs}^{\rm max}$,  $E_{\rm BL}$ and 
($E_{\rm BL}+E_{\rm D}$) by the
relations $E_{\rm BL} = L_{\rm BL}/\dot {M}$, 
$E_{\rm BL}+E_{\rm D} = (L_{\rm BL}+L_{\rm D})/\dot {M}$ and 
%$T_{\rm obs}^{\rm max}=T_{\rm col}^{\rm max}/(f {\dot {M}}^{1/4})$ 
$T_{\rm obs}^{\rm max}=T_{\rm col}^{\rm max}/f$. 
%(because here  
%$T_{\rm obs}^{\rm max}$ is in the unit of ${\dot {M}}^{1/4}$).
By interpolation, we calculate two corresponding 
limiting $\Omega_{\rm *}$'s (i.e., the allowed range in $\Omega_{\rm *}$)
for each fitted quantity. We take the common
region of these three ranges, which is the net allowed
range in $\Omega_{\rm *}$. We do this for $\dot {M}$'s in the
range $10^{-13}~M_{\odot} y^{-1}$ to $10^{-6}~M_{\odot} y^{-1}$ 
(which is reasonable for LMXB's) with logarithmic interval
0.0001, for a particular value of $f$.  If for some $\dot {M}$,
there is no allowed $\Omega_{\rm *}$, then that value of $\dot {M}$ is not 
allowed.  Thus we get the allowed range of $\dot {M}$ for a particular $f$. 
Next we repeat the whole procedure described above for various values
of $f$, in the range 1 to 10. If for some $f$, there is no
allowed $\dot {M}$, then that $f$ is not allowed. Thus we get
an allowed range of $f$. Taking the union of all the allowed
ranges of $\dot {M}$, we get the net allowed range of $\dot {M}$ 
(and similarly the net allowed range of $\Omega_{\rm *}$) for
a particular EOS, gravitational mass and a set of error bars. 
The allowed range of $E_{\rm BL}$, $E_{\rm D}$, $R$, etc. then
easily follow, since their general variations with respect to 
$\Omega_{\ast}$ are already known.  The results of this exercise 
for various equations of state is shown in Table 2.  From Table 2 we
can read off the allowed range in $f$, $\nu_{\ast}=\Omega_{\ast}/2\pi$,
$\nu_{\rm in}=\Omega_{\rm K, in}/2\pi$, $R$, $r_{\rm eff}^{\rm max}$,
$\dot{M}$. For e.g. for gravitational mass $M=1.4$~\msun, an assumed
uncertainty of $25$\% in the luminosity, and $10$\% uncertainty in color
temperature, these are respectively ($1.37$--$2.39$), ($0.736$--$1.755$)~kHz, 
($0.745$--$1.787$)~kHz, ($11.3$--$20.7$)~km, ($16.0$--$28.3$)~km,
($11.2$--$34.6$)~$\dot{M}_{\rm Edd}$.  On relaxing the conditions on 
luminosity and color temperature to $50$\% and $20$\% respectively,
the corresponding ranges change to ($1.16$--$2.97$), ($0.719$--$1.743$)~kHz, 
($0.742$--$1.755$)~kHz, ($10.7$--$20.7$)~km, ($15.6$--$28.4$)~km, 
($5.8$--$42.4$)~$\dot{M}_{\rm Edd}$.

The EOS model (A) is the softest in the sample. The maximum
mass of neutron stars (at $\Omega_{\ast}=\Omega_{\rm ms}$) corresponding
to this EOS is 1.63 $M_\odot$.  So the constraint results for Cygnus X--2,
using this EOS are done only for $M=1.4$~\msun.
Since the luminosity in the boundary layer is
about 10\% of the disk luminosity, the neutron star is expected to
be spinning close to the maximum allowed value. This is reflected 
in our results by the ratio of $\Omega_{\ast}/\Omega_{\rm ms} \approx 0.95$.
In all these cases, the neutron star radius happens to be larger than
the innermost stable circular orbit. Hence the radius of the inner edge of 
the disk coincides with
the neutron star radius. Therefore, the angular velocity of the particles
in the disk inner edge will be very nearly equal to that of the neutron 
star. This implies that the viscous torque in the disk inner edge will 
not be very significant, and the use of the Page \& Thorne (1974) formalism
will not introduce any gross error in the constraint estimates presented 
by us. 

%% COMMENT ON THE RESULTS OF f
\section{Summary and Discussion}\label{sec: V}

In this paper, we have calculated the temperature profiles of accretion
disks around rapidly rotating and non--magnetized neutron stars, using
a fully general relativistic formalism.  The maximum temperature
and its location in the disk are found to differ substantially from their
values corresponding to the Schwarzschild space--time, depending on the 
rotation rate of the accreting neutron star.  We have considered a model
for the spectrum of the X--ray emission from the accretion disk, 
parameterized by the mass accretion rate, the color temperature, 
and the rotation rate of the star.  We have compared the maximum effective
temperature in the disk and the accretion luminosities (corresponding to
the disk and the boundary layer) with the results of spectral fitting 
for the X--ray source Cygnus X--2 (WSP), and
derived constraints on these parameters for the neutron star in this
X--ray binary.  The main conclusion of our analysis is that the neutron 
star in Cygnus X--2 has a rapid spin rate (close to the centrifugal mass
shed value), and that the system has a fairly large accretion rate 
(several times $10^{18}$~g~s$^{-1}$).  The low luminosity of 
the boundary layer compared to that of the disk for Cygnus X--2 
is consistent with the above conclusion that the neutron star in this
system has a rapid rotation rate. 
The low value of the ratio 
$L_{\rm BL}/L_{\rm D}$ justifies our 
assumption that the radiation pressure is negligible in the disk, so that 
the geometrically thin approximation for the disk is reasonable.
According to Shimura \& Takahara (1988), the spectrum from the disk can 
be represented as a multi-color blackbody only if 
$\dot{M} >0.1 \dot{M}_{\rm edd}$. 
Our results for Cygnus X--2 are in accord with this.
Interestingly, if we take the lower value $1.7$ for the color index $f$ 
(Shimura \& Takahara 1988), we obtain a consistent set of results,
except for the stiffest EOS model (D).  This suggests that the comparatively
lower values of $f$ would disfavor stiff EOS for neutron star matter.
However, if we take the value of $f=2.6$, as reported by Borozdin et al. 
(1999), one would require an EOS model that is stiffer
than the stiffest used here, or a mass greater than $M=1.78$~\msun (if one
uses the narrower limits on the luminosity and color temperature).  On the 
other hand, if one were to use the broader limits, the hardening factor 
$f=2.6$ is disallowed only by the softest EOS model.

We have assumed here that the magnetic field of the neutron star is weak, 
which implies that the radius of the last orbit of the accretion disk
should be much greater than the Alfv\'{e}n radius ($r_{\rm a}$) 
(e.g., Shapiro \& Teukolsky 1983),
\begin{equation}
R >> r_{\rm a} = 2.9 \times 10^7  {({\dot {M}\over \dot 
{M}_{\rm edd}})}^{-2/7} \mu_{30}^{4/7} ({M\over M_\odot})^{-3/7} 
\end{equation}
\noindent where 
$M$ is the mass of the neutron star, $\mu_{30}$ is
the magnetic moment in units of $10^{30}$ G cm$^3$ and $r_{\rm a}$ is in cm.
 The above condition implies
that for $R \approx 15$ km., $\dot{M}/\dot {M}_{\rm edd} \approx 20$ and 
$M = 1.4 M_\odot$, the magnetic moment  $\mu_{30} << 3.4 \times 10^{-2} $ 
or the magnetic field in the surface should be less than $10^{10}$ G.
So the conclusions presented by us will be valid for the neutron star 
magnetic field upto a few times $10^{9}$~G.

In our analysis, we have assumed that the the boundary layer between the
disk and the neutron star surface does not affect the inner regions of
the disk. This will be a valid approximation when the boundary layer 
luminosity is smaller than the disk luminosity, 
and the boundary layer extent is small compared to the radius
of the star. The flux received at earth  from this region is

\begin{equation}
F_{\rm BL} = ({2 \pi R \frac{\Delta R}{D^2}}) \cos{\theta} 
({\sigma T_{\rm BL}^4 \over \pi})
\end{equation}

\noindent where $\Delta R$ is the width of the boundary layer, $D = 8$ kpc 
is the distance to the source, $\theta = 60^o$ is the inclination angle and
$T_{\rm BL}$ is the effective temperature. Spectral fitting gives 
a best fit value
for $F_{\rm BL} \approx 4 \times 10^{-9}$ ergs sec$^{-1}$ cm$^{-2}$ 
and $T_{\rm BL} = T_{\rm col(BL)}/f_{\rm BL}$ = 2.88/$f_{\rm BL}$~keV, 
where $f_{\rm BL}$ is the color factor for
the boundary layer and $T_{\rm col(BL)}$ is the color temperature of the
boundary layer. Using these values, $\Delta R \approx $ 0.2 
$f_{\rm BL}^4$ km, 
which is indeed smaller than $R$ provided the boundary layer color 
factor $f_{\rm BL}$ is close to unity. This is supported 
by the work of London, Taam \& Howard (1986) 
and Ebisuzaki (1987), who obtain $f_{\rm BL}\approx 1.5$.

A few comments regarding the validity of the Page \& Thorne (1974) 
formalism for accreting neutron star binaries are in order here.
%Unlike the black hole binary case, in neutron star binaries the 
%disk inner edge may coincide with the neutron star surface.  
Unlike for the case of black holes, neutron stars possess hard
surface that could be located outside the marginally stable orbit.
For neutron star binaries, this gives rise to a possiblity of 
the disk inner edge coinciding with
the neutron star surface.  We
have assumed that the torque (and hence the flux of energy) vanishes
at the disk inner edge even in cases where the latter touches
the neutron star surface.  In the case of rapid spin of the neutron 
star (as we infer for Cygnus X--2), the angular velocity of a particle in 
Keplerian orbit at disk 
inner edge will be close to the rotation rate of the neutron star.
Therefore, the torque between the neutron star surface and the inner
edge of the disk is expected to be negligible.  Independently of 
whether or not the neutron star spin is large, Page \& Thorne (1974)
argued that the error in the calculation of $T_{\rm eff}$ will not be 
substantial outside a radial
 distance $r_{\rm o}$, where $r_{\rm o}$ is given by $r_{\rm o} - r_{\rm in} =
0.1 r_{\rm in}$. In our calculation, we find that $r_{\rm eff}^{\rm max}$ 
(which is the most important region for the generation of X--rays) is
greater than $r_{\rm o}$ by several kilometers for all the cases
considered.

The shortest time-scale of the system is given by the frequency in the 
innermost stable circular orbit ($\nu_{\rm in}$, Table 2: column 5). A 
periodic oscillation in the 
system should be at a frequency lower than $\nu_{\rm in}$ (unless 
the model invoked to explain the temporal behavior predicts substantial
power in the second harmonic, i.e.,  $\nu_{\rm QPO} \approx 2 \nu_{\rm in}$). 
The maximum frequency of the
kHz quasi-periodic oscillation (QPO) observed for Cygnus X-2 is 1005 Hz
(Wijnands et al. 1998). The stiffest EOS, model (D), will then be 
disfavored since $\nu_{\rm in} < 1$~kHz for this model. 
Further, the neutron star mass estimate in Cygnus X-2 ($ \approx  1.78 
M_\odot$, Orosz \& Kuulkers 1998) is not consistent with the soft 
EOS model (A). Our analysis, therefore, favors neutron star EOS model 
which are intermediate in the stiffness parameters.

We have not attempted to model the observed temporal behavior of 
the source, and in particular, the QPO observations. Beat frequency models
identify the peak separation of the two kHz QPO observed with the neutron 
star spin rate. 
For Cygnus X-2 the observed peak separation is $\Delta \nu = 346 
\pm 29$ Hz (Wijnands et al. 1998) which is smaller than the typical rotation 
frequencies
calculated here. However, a pure beat-frequency model has been called into
question due to several observations. For instance, $\Delta \nu$ has been 
observed to vary by about
40\% for Sco X-1 (van der Klis et al. 1997) and the kHz QPO frequencies 
have been found to be correlated with the  break frequency
($\approx$ 20 Hz) of the power spectrum density.  An alternate model,
where the QPOs are suggested to originate due to non--Keplerian motion 
of matter in the disk (Titarchuk \& Osherovich 1999; Osherovich \& 
Titarchuk 1999a; Osherovich \& Titarchuk 1999b; Titarchuk, Osherovich \& 
Kuznetsov 1999) have been proposed.  These authors have also demonstrated
the model by applying it to particular sources.  Inclusion of this Newtonian 
model into the framework of the calculations mentioned in this paper require
a parallel formulation within the space--time geometry chosen herein. 

X-ray binaries like Cygnus X-2 are believed to be the progenitors of
the millisecond pulsars. Therefore, the discovery of a pulsar with a 
period $\approx 1 $ ms will strengthen the model presented in this 
paper, in terms of a rapidly rotating accreting neutron star. X-ray 
spectral analysis of Cygnus X-2 and similar sources using data from recent
satellites (e.g. BeppoSAX, ASCA, Chandra) are required to provide further 
support to the model presented in this paper.

\acknowledgements

We thank Paul J. Wiita for reading the manuscript and 
suggesting improvements in the presentation.

\input psbox.tex
\newpage
\begin{table*}
\begin{center}
%\caption{Centrifugal mass shed limit ($\Omega_{\rm ms}$), the neutron star
\caption{Theoretically computed parameters: Centrifugal mass shed limit 
($\Omega_{\rm ms}$), the neutron star
radius ($R$), the disk inner edge radius ($r_{\rm in}$), specific gravitational 
energy release in the boundary layer ($E_{\rm BL}$) and in the disk 
($E_{\rm D}$), their ratio $E_{\rm BL}/E_{\rm D}$, the maximum effective 
temperature ($T_{\rm eff}^{\rm max}$),
the radial location ($r_{\rm eff}^{\rm max}$) in the disk corresponding to 
$T_{\rm eff}^{\rm max}$, $T_{\rm obs}^{\rm max}$ (see text) and the 
radial location ($r_{\rm obs}^{\rm max}$) corresponding to this.  These
values are listed for two values of $M$ for all EOS models considered 
here (except for EOS model (A), where the maximum neutron star mass is less than 
$1.78$~\msun, so only $M=1.4$~\msun is considered). The number following
the letter $E$ represents powers of $10$.}
\vspace{-3.0cm}
\hspace{-3.0cm}
{\mbox{\psboxto(19cm;29.5cm){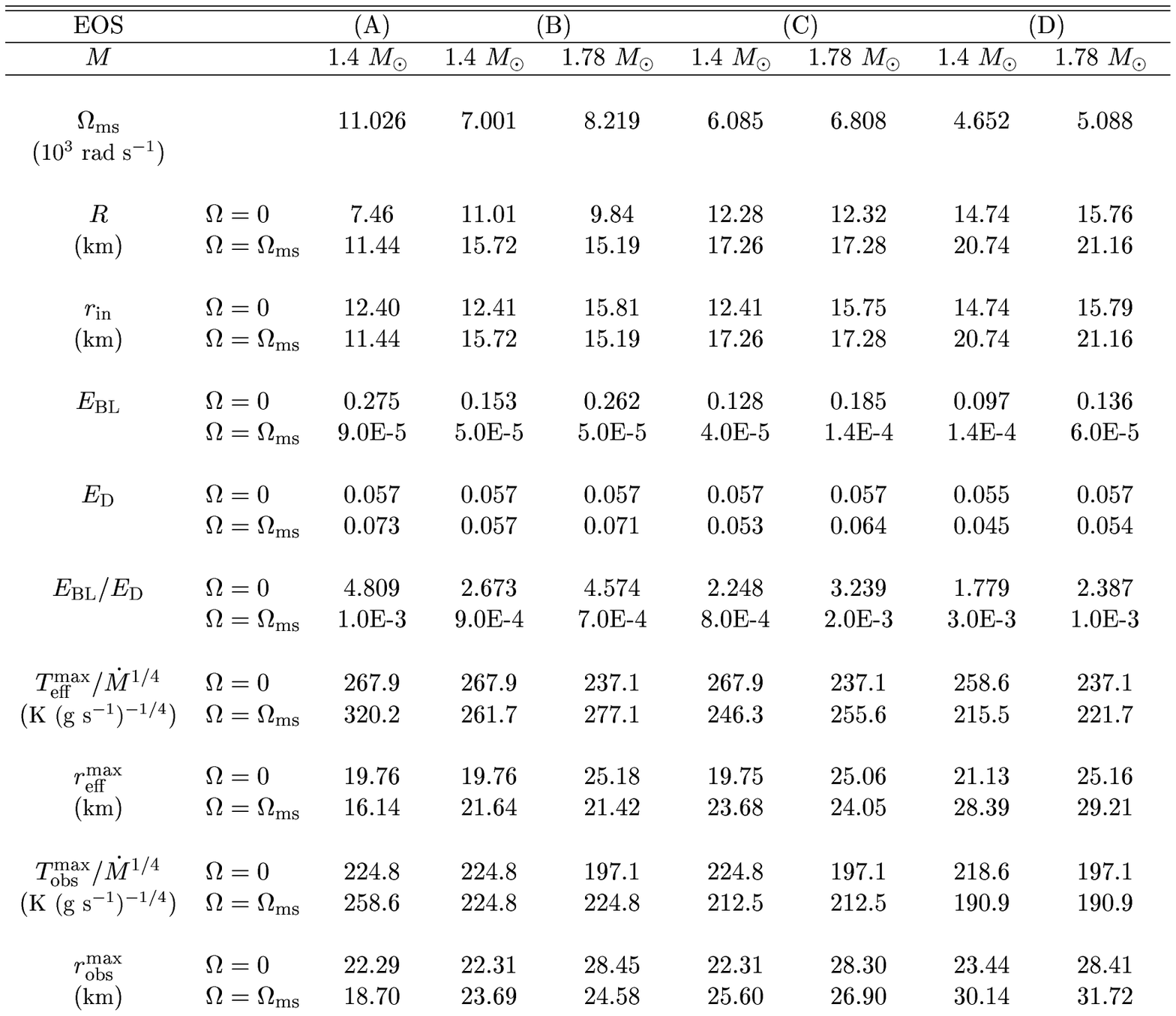}}}
\end{center}
\end{table*}
\newpage
\begin{table*}
%\caption{Allowed range of  parameters for various EOS : (A),
\caption{Observational constraints for various EOS : (A),
(B), (C), (D).  L and U stand for lower and upper limits. The parameters 
are f (color factor), $\nu_{\ast}$ (frequency of the neutron star), 
$\nu_{\rm in}$ (frequency of the last orbit in the disk),
$R$ ( radius of the neutron star), $r_{\rm eff}^{\rm max}$ 
(radius where the effective temperature of the disk is maximum) and 
$\dot {M}$ (the accretion rate).  The limits are for 25\% uncertainty in 
luminosity and 10\% uncertainty in the color temperature.
Values in [ ] are for 50\% uncertainty in luminosity and 20\% uncertainty 
in the color temperature.  For EOS (A), the mass of the neutron star cannot 
exceed 1.63 $M_\odot$ hence the $1.78 M_\odot$ solution
is not presented. $\dot {M}_{\rm edd}$ is the Eddington accretion rate, which
is $1.4\times 10^{17} M/$\msun~g~s$^{-1}$, where $M$ is the neutron star 
mass.}
\begin{tabular}{lllllllll}
\hline
\hline
EOS  & $M$ &  & $f$  & $\nu_{\ast}$& $\nu_{\rm in}$ & $R$ & 
$r_{\rm eff}^{\rm max}$ & $\dot{M}$   \\
   & $M_\odot$ &  & &  kHz & kHz & km & km & $\dot{M}_{\rm edd}$  \\
\hline
(A)  &   1.4 & L  & 1.37[1.16] & 1.753[1.743] & 1.755[1.755] & 11.3[10.7] & 16.0[15.6] & 11.2[5.8]  \\
  &  &       U  & 1.99[2.56] & 1.755[1.755] & 1.787[1.944] & 11.4[11.4] & 16.1[16.1] & 22.9[27.5]  \\
\hline
(B)  &   1.4 & L  & 1.53[1.29] & 1.106[1.087] & 1.132[1.123] & 15.2[14.3] & 21.0[20.0] & 13.8[7.2] \\
  &  &       U  & 2.18[2.74] & 1.112[1.113] & 1.177[1.285] & 15.6[15.6] & 21.5[21.6] & 27.0[33.5]  \\
\hline
(C)  &   1.4 & L  & 1.57[1.33] & 0.964[0.945] & 0.975[0.971] & 16.8[15.6] & 23.1[21.7] & 14.9[7.7]  \\
  &  &       U  & 2.24[2.81] & 0.968[0.968] & 1.015[1.134] & 17.2[17.2] & 23.6[23.7] & 29.3[36.5]  \\
\hline
(D)  &   1.4 & L  & 1.67[1.42] & 0.736[0.719] & 0.745[0.742] & 20.1[18.6] & 27.6[25.7] & 17.5[9.1]  \\
  &  &       U  & 2.38[2.97] & 0.740[0.740] & 0.779[0.876] & 20.7[20.7] & 28.3[28.4] & 34.6[42.4]  \\
\hline
(B)  &   1.78 & L  & 1.58[1.33] & 1.303[1.292] & 1.322[1.315] & 14.8[14.2] & 21.2[20.7] & 8.9[4.7]  \\
  &  &        U  & 2.28[2.91] & 1.307[1.307] & 1.361[1.462] & 15.1[15.1] & 21.4[21.4] & 17.2[21.4]  \\
\hline
(C)  &   1.78 & L  & 1.65[1.39] & 1.081[1.067] & 1.086[1.085] & 17.1[16.2] & 23.8[23.0] & 9.8[5.1]  \\
  &  &        U  & 2.39[3.01] & 1.083[1.083] & 1.109[1.209] & 17.3[17.3] & 24.0[24.1] & 19.3[24.0]  \\
\hline
(D)  &   1.78 & L  & 1.74[1.47]  & 0.806[0.791] & 0.817[0.813] & 20.6[19.2] & 28.6[27.1] & 11.4[6.0]  \\
  &  &        U  & 2.50[3.15]  & 0.809[0.809] & 0.848[0.938] & 21.1[21.1] & 29.1[29.2] & 22.2[27.7]  \\
\hline
\end{tabular}

\end{table*}

\newpage
\input psbox.tex
\begin{figure*}[h]
\hspace{-1.5cm}
{\mbox{\psboxto(17cm;20cm){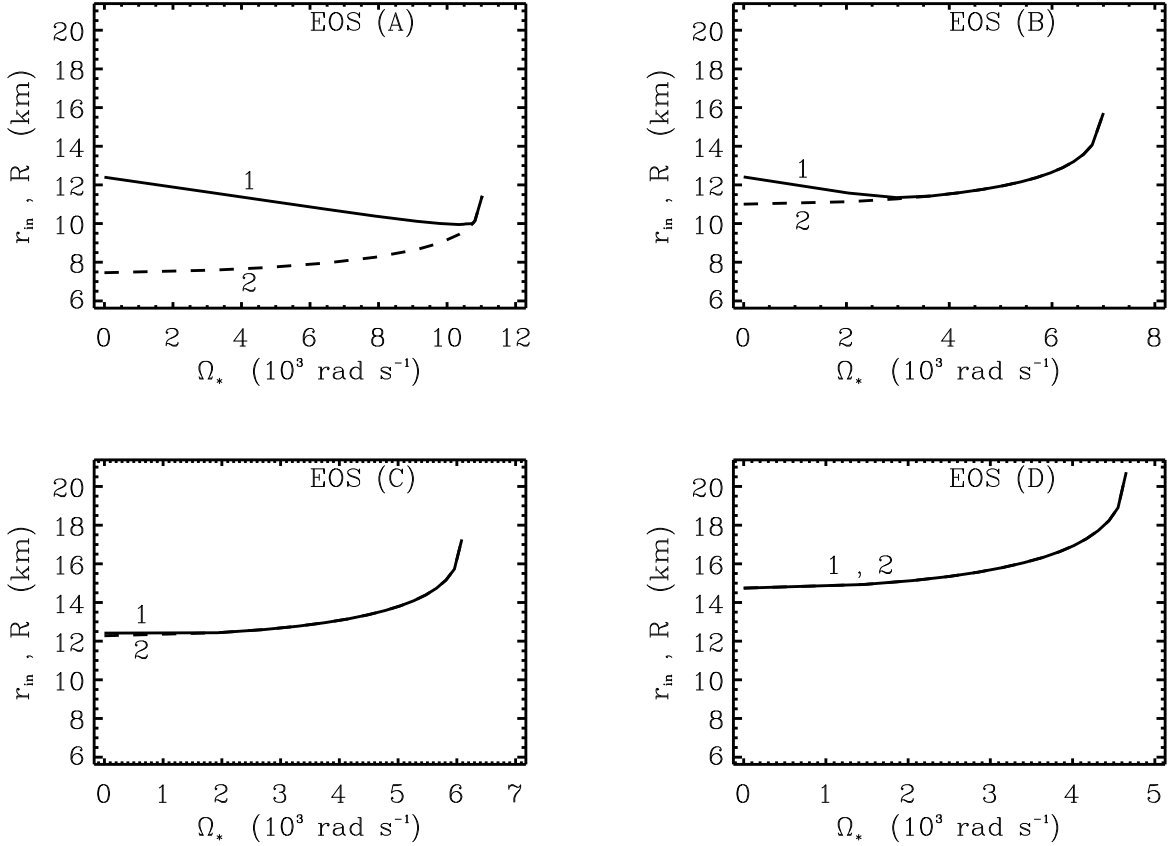}}}
\caption{Disk inner edge radius ($r_{\rm in}$, curve 1) and neutron star 
radius ($R$, curve 2),
as functions of neutron star angular velocity ($\Omega_{\ast}$) for
various EOS models.  The curves are for a fixed gravitational mass 
($M=1.4$~\msun) of the neutron star.}
\end{figure*}

\newpage
\begin{figure*}[h]
\hspace{-1.5cm}
{\mbox{\psboxto(17cm;20cm){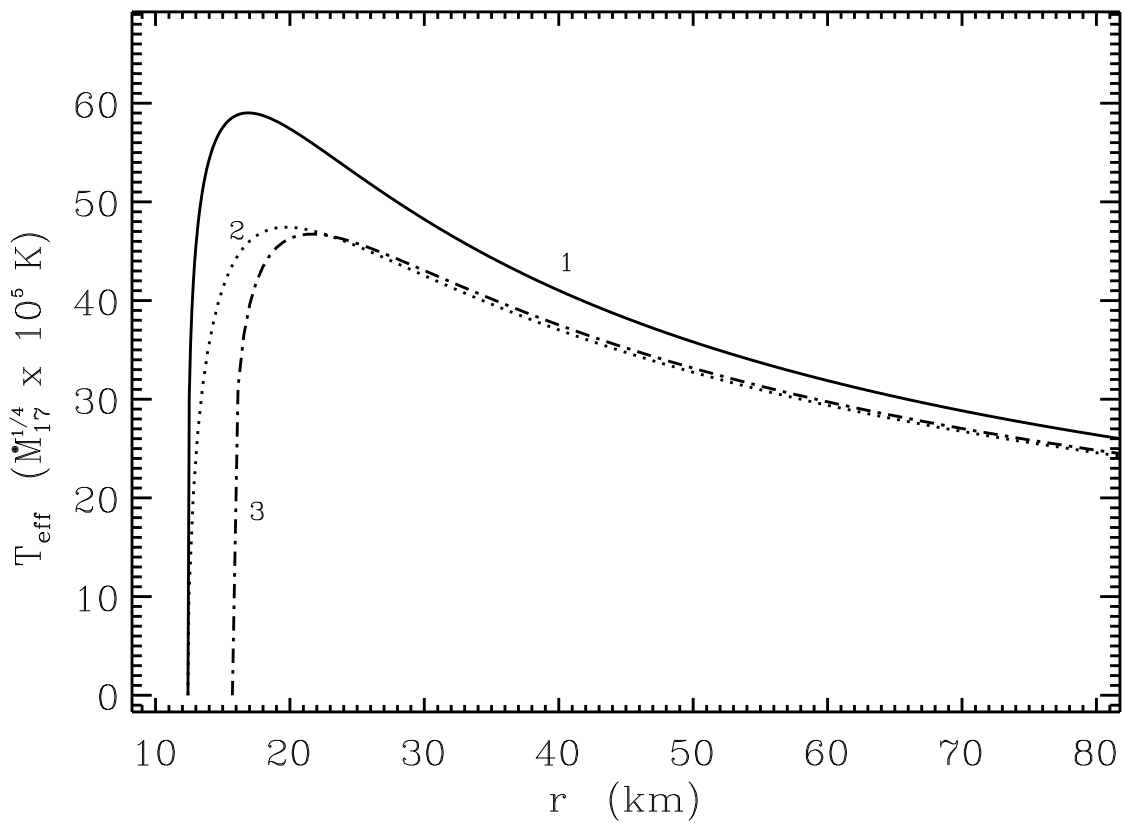}}}
\caption{General relativistic corrections to Newtonian temperature profiles 
for EOS model (B) and neutron star gravitational mass $M=1.4$~\msun. 
Curve (1) corresponds to the Newtonian case, curve (2) to the 
Schwarzschild case and curve (3) to a neutron star rotating at
the centrifugal mass shed limit, calculated using the 
metric ~(\ref{eq: metric}). For curve (1) it is assumed that, 
$r_{\rm in}=6 G M/c^2$.
In this and all subsequent figures, the temperature is expressed in units 
of $\dot{M}_{17}^{1/4}~10^5$~K, where $\dot{M}_{17}$ is the steady state mass 
accretion rate in units of $10^{17}$~g~s$^{-1}$.}
\end{figure*}

\newpage
\begin{figure*}[h]
\hspace{-1.5cm} 
{\mbox{\psboxto(17cm;20cm){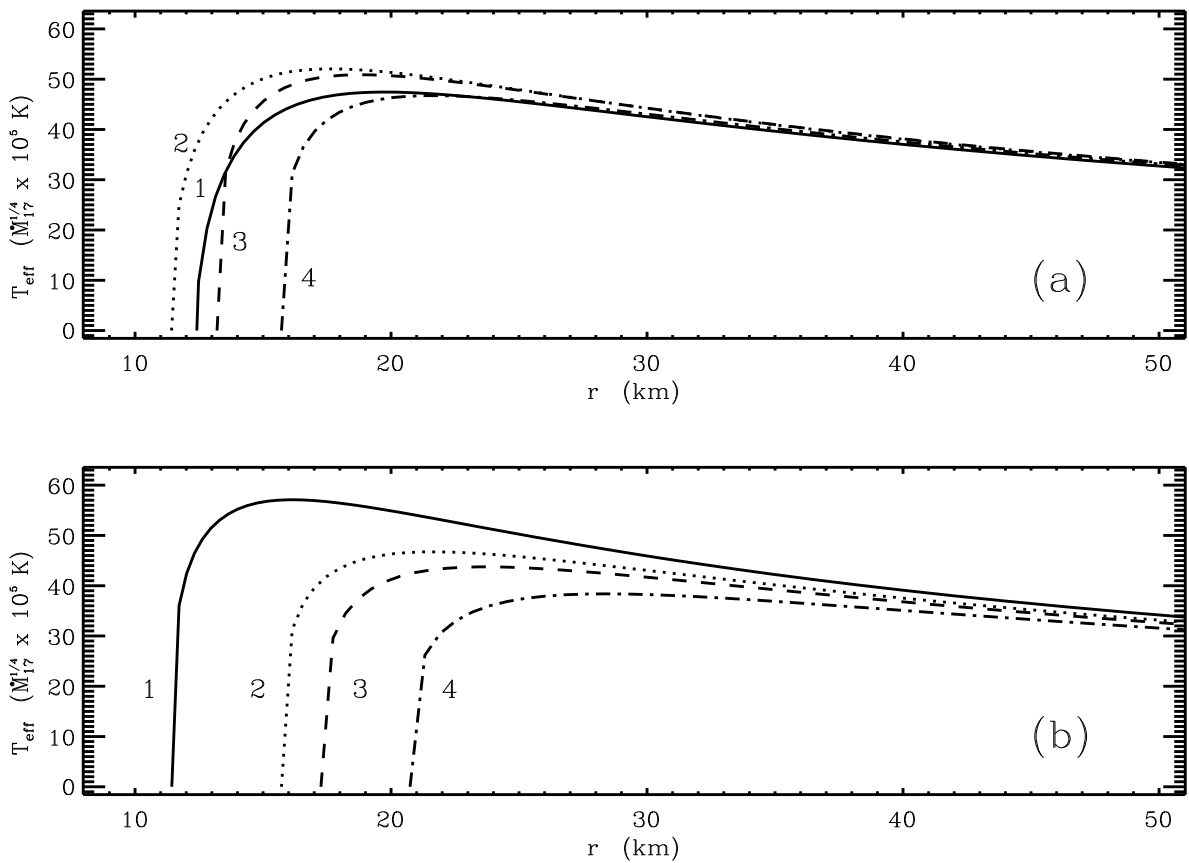}}}
\caption{Temperature profiles incorporating the effects of rotation of the
neutron star. The plots correspond to (a) EOS model (B) and an assumed neutron 
star mass of $M=1.4$~\msun for rotation rates: $\Omega_{\ast}=0$ (curve 1), 
$\Omega_{\ast}=3.647\times10^3$~rad~s$^{-1}$ (curve 2), 
$\Omega_{\ast}=6.420\times10^3$~rad~s$^{-1}$ (curve 3), 
$\Omega_{\ast}=7.001\times10^3$~rad~s$^{-1}= \Omega_{\rm ms}$ (curve 4) 
(b) the same assumed mass and $\Omega_{\ast}=\Omega_{\rm ms}$ for
the four EOS models (A):curve 1, (B):curve 2, (C):curve 3 and (D):curve 4.} 
\end{figure*}

\newpage
\begin{figure*}[h]
\hspace{-1.5cm}
{\mbox{\psboxto(17cm;20cm){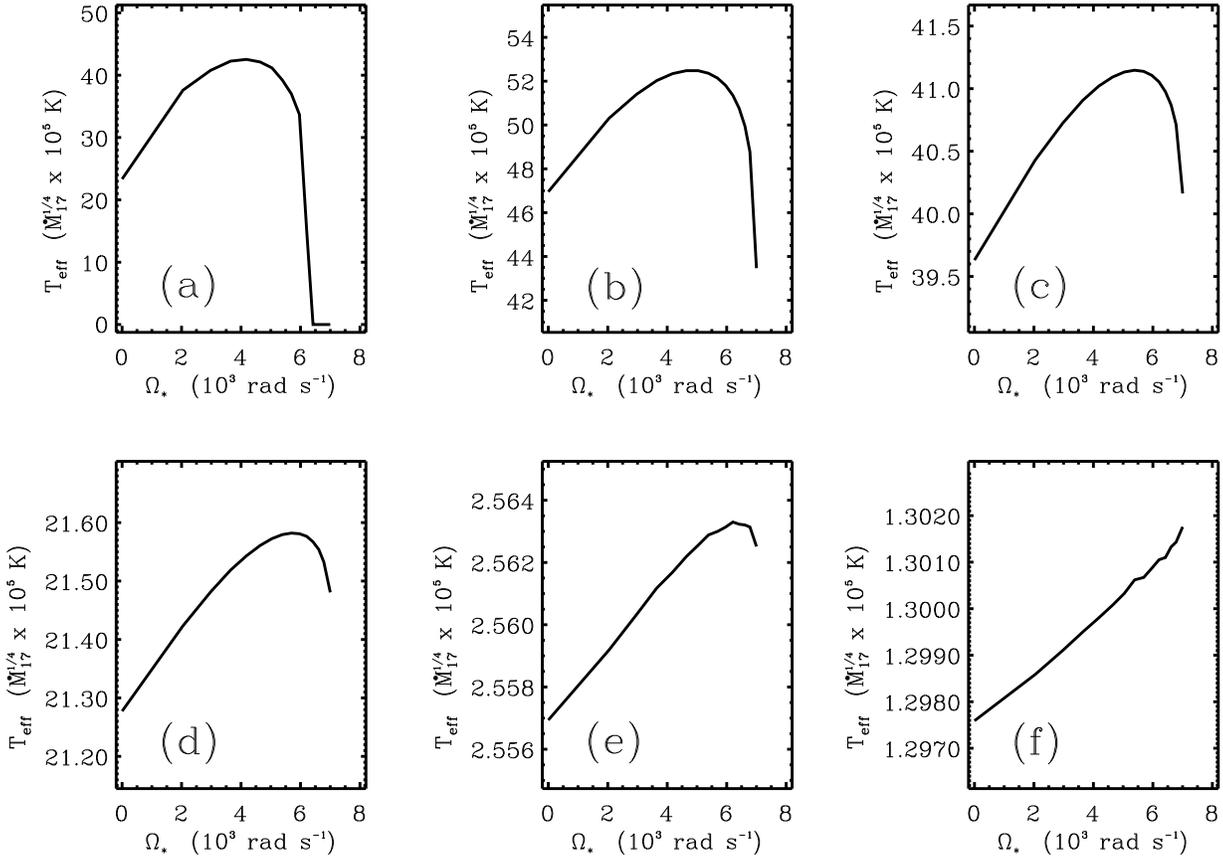}}}
\caption{Plot of $T_{\rm eff}$ versus $\Omega_{\ast}$ for
chosen constant radial distances for fixed neutron star mass $M=1.4$~\msun 
and EOS (B). The plots correspond to:
(a) $r=13$~km, (b) $r=18$~km, (c) $r=35$~km, (d) $r=100$~km, (e) $r=2000$~km, 
(f) $r=5000$~km.}
\end{figure*}

\newpage
\begin{figure*}[h]
\hspace{-1.5cm}
{\mbox{\psboxto(17cm;20cm){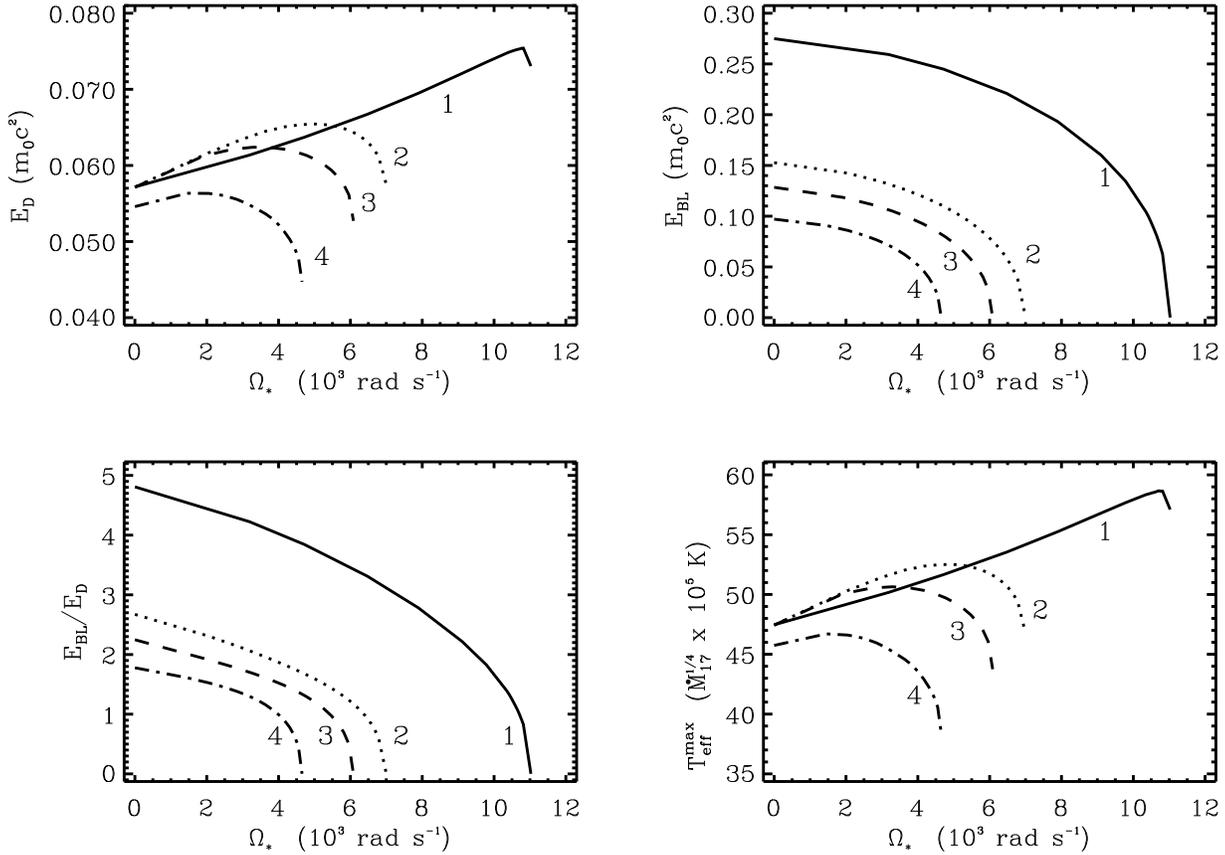}}}
\caption{The variations of the $E_{\rm D}$, $E_{\rm BL}$, 
$E_{\rm BL}/E_{\rm D}$ and $T_{\rm eff}^{\rm max}$ with
$\Omega_{\ast}$ for a chosen neutron star mass value of $1.4$~\msun for 
the four EOS models. The curves have the same significance as Fig. 3b.}
\end{figure*}

\newpage
\begin{figure*}[h]
\hspace{-1.5cm}
{\mbox{\psboxto(17cm;20cm){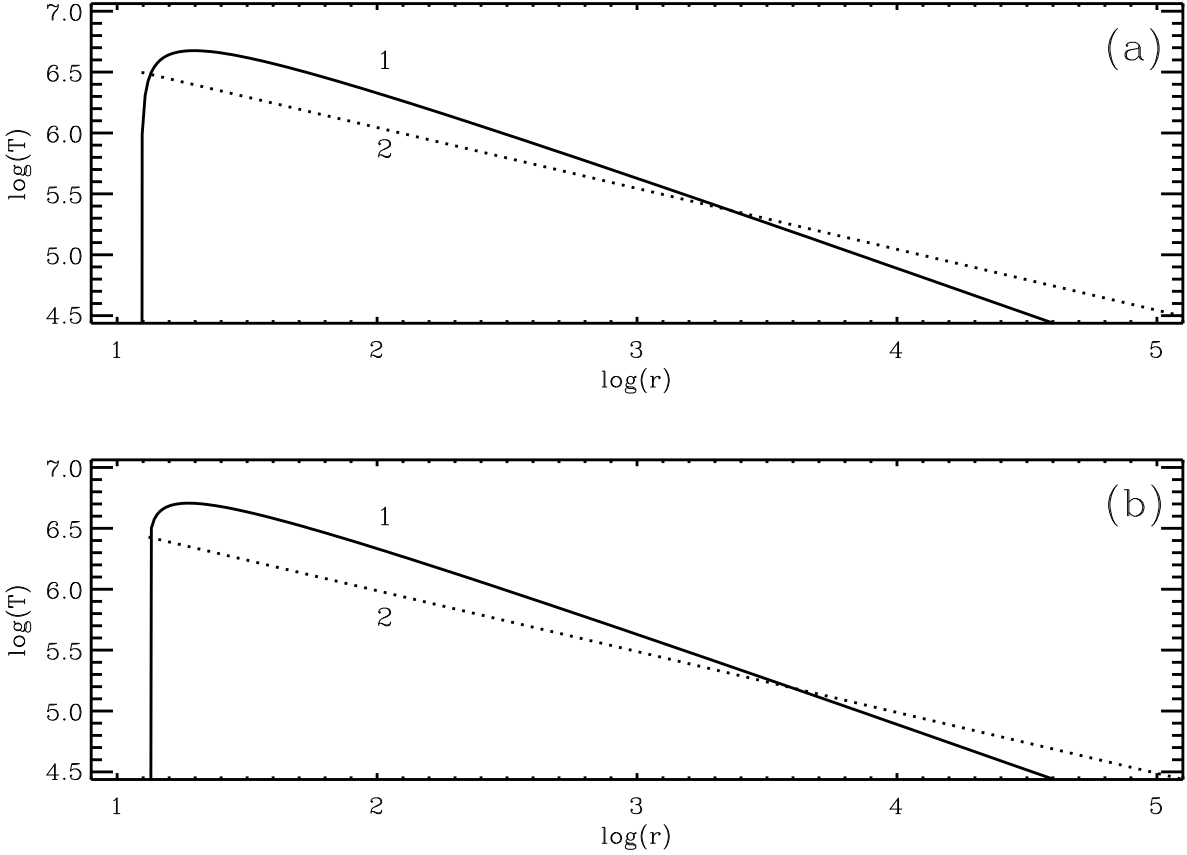}}}
\caption{Comparison between the radial profiles of $T_{\rm eff}$ (curve 1) 
and $T_{\rm irr}$ (curve 2), calculated for $\eta=E_{\rm BL}+E_{\rm D}$, 
$\beta=0.9$, 
$h/r=0.2$ and $n=9/7$ in Eq. (\ref{eq: tirr}),
for two values of neutron star spin rates: 
(a) $\Omega_{\ast}=0$ and (b) $\Omega_{\ast}=6.420\times10^{3}$~rad~s$^{-1}$.
The curves are for a neutron star configuration having 
$M=1.4$~\msun, described by EOS model (B). The temperatures are in
units of $\dot{M}_{17}^{1/4}$ and radial extent in km.  For 
illustrative purposes, we have displayed this comparison in a log-log plot.}
\end{figure*}

\newpage
\begin{figure*}[h]
\hspace{-1.5cm}
{\mbox{\psboxto(17cm;20cm){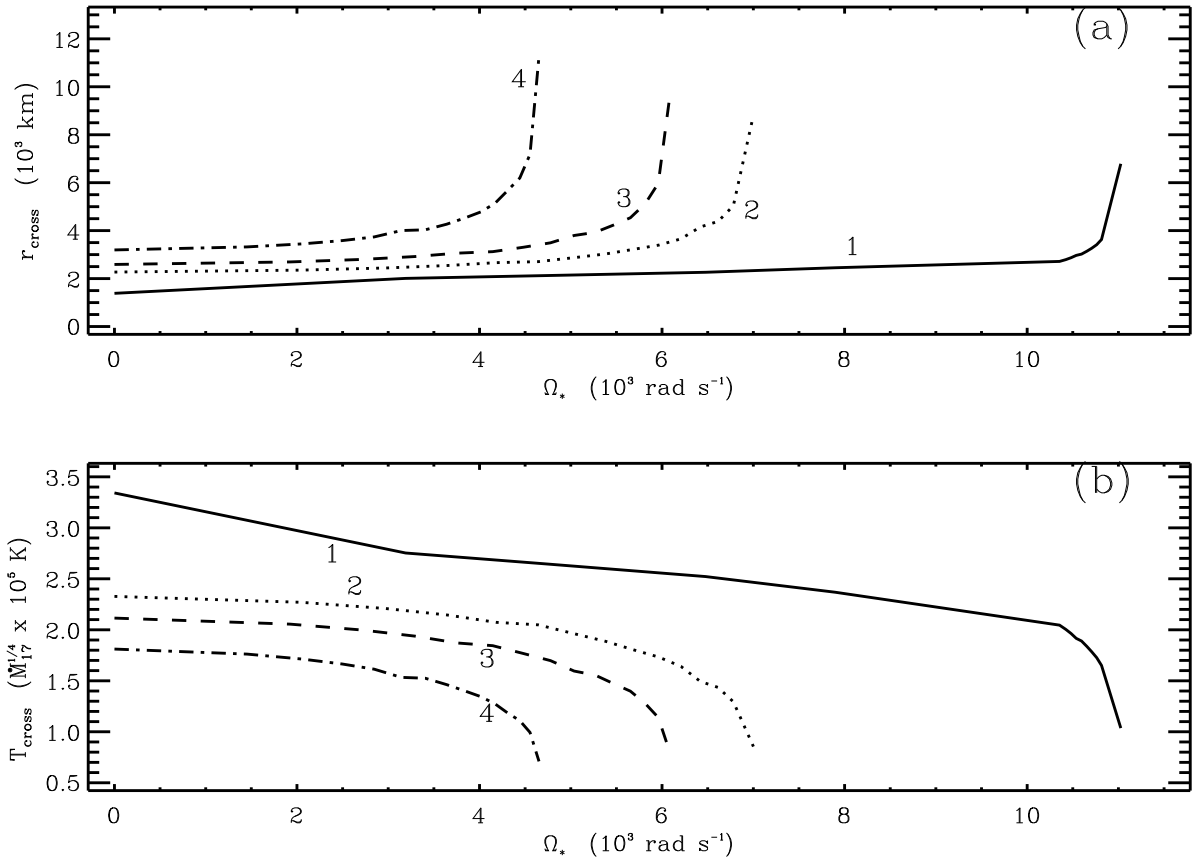}}}
\caption{Plots for: (a) $r_{\rm cross}$ versus $\Omega_{\ast}$ 
and (b) $T_{\rm cross}$ versus $\Omega_{\ast}$.  These 
are for fixed neutron star gravitational mass of $M=1.4$~\msun and for
the different EOS models as in Fig. 3b. Here $T_{\rm irr}$ is  calculated
for $\eta=E_{\rm BL}+E_{\rm D}$, $\beta=0.9$, $h/r=0.2$ and $n=9/7$.} 
\end{figure*}

\newpage
\begin{figure*}[h]
\hspace{-1.5cm}
{\mbox{\psboxto(17cm;20cm){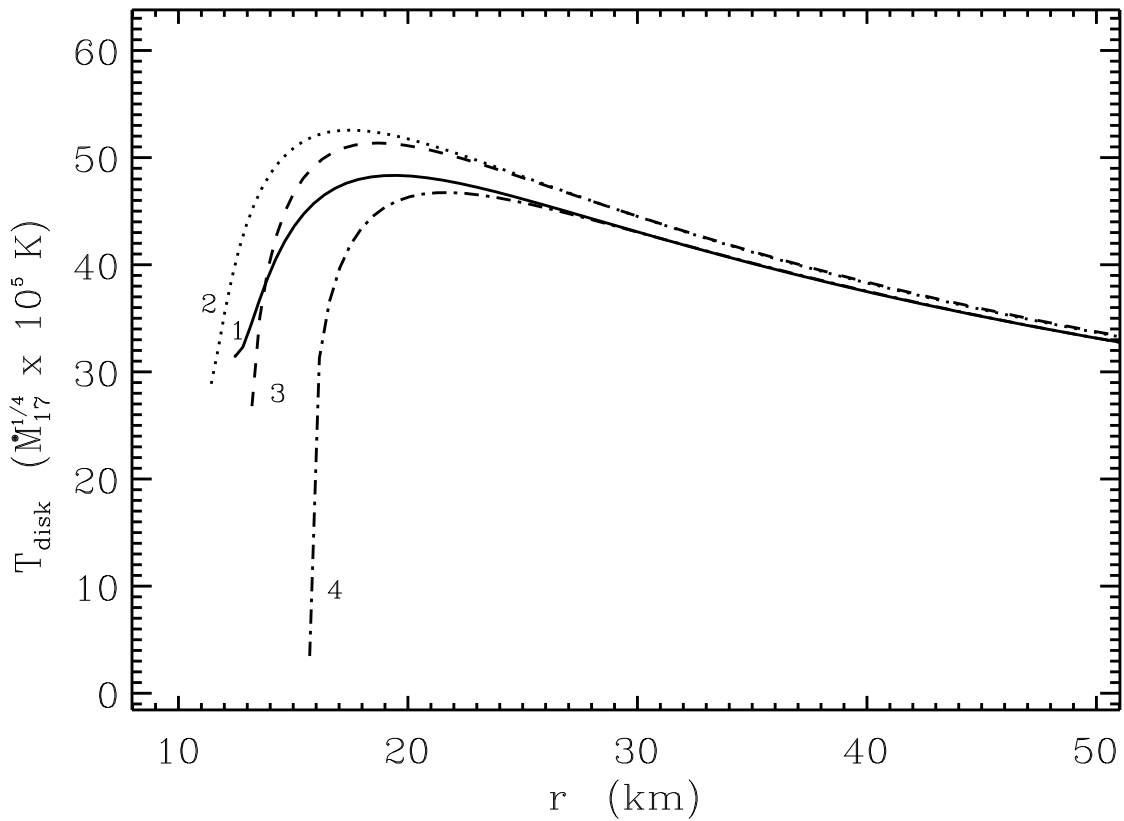}}}
\caption{The disk temperature ($T_{\rm disk}$) profiles for
a $M=1.4$~\msun neutron star corresponding to EOS model (B) having various 
rotation rates as in Fig. 3a. These curves are obtained for $\eta=E_{\rm BL}$,
and the same values of  $\beta$, $h/r$ and $n$ as in Fig. 6.}
\end{figure*}

\newpage
\begin{figure*}[h]
\hspace{-1.5cm}
{\mbox{\psboxto(17cm;20cm){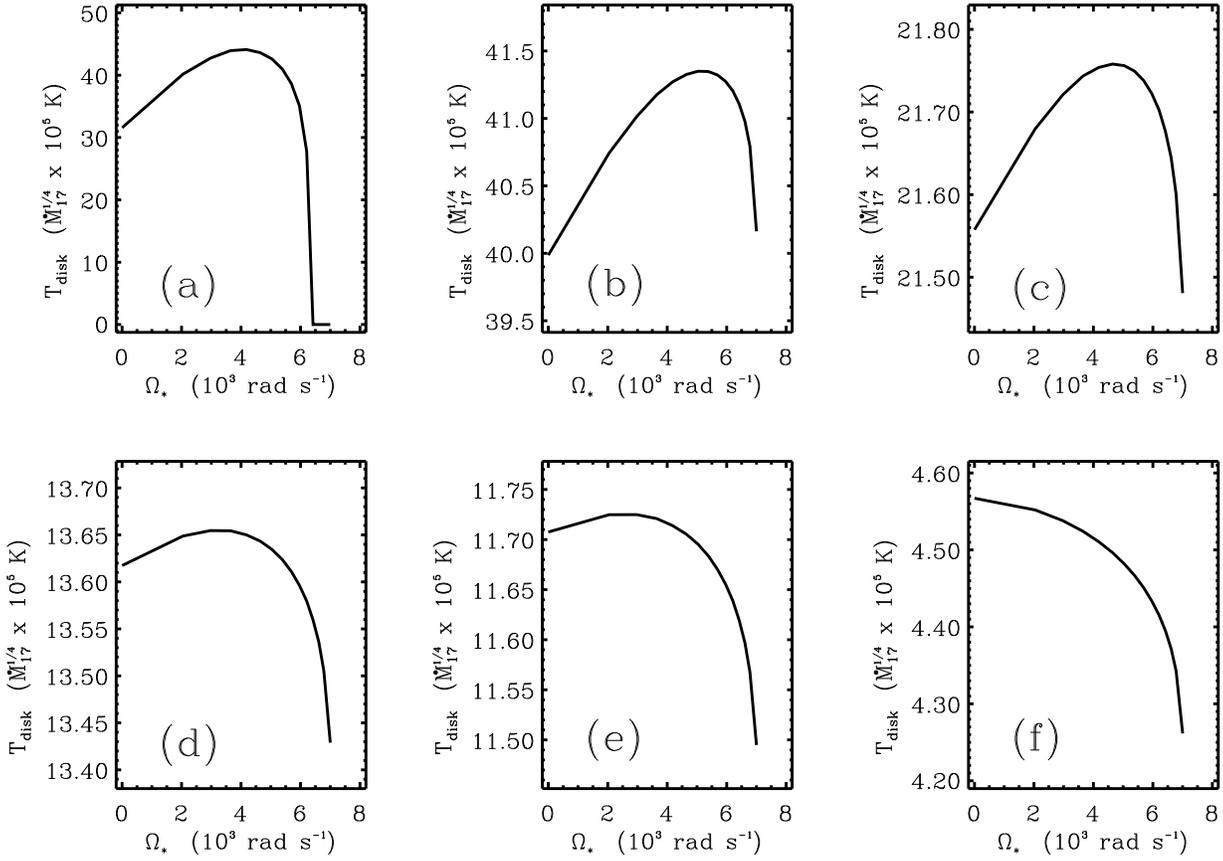}}}
\caption{Plots of $T_{\rm disk}$ versus $\Omega_{\ast}$ at 
various chosen radial distances: (a) $r= 13$~km, (b) $r= 35$~km,
(c) $r= 100$~km, (d) $r= 200$~km, (e) $r= 250$~km, (f) $r= 1000$~km.
These are for EOS model (B), an assumed gravitational mass value of 
$1.4$~\msun, and the same 
values of $\eta$, $\beta$, $h/r$ and $n$ as in Fig. 7.} 
\end{figure*}

\end{document}